# Measuring and estimating interaction between exposures on dichotomous outcome of a population

**Xiaoqin Wang[1], Weimin Ye[2] and Li Yin[2,*]**

[1]Department of Electronics, Mathematics and Natural Sciences, University of Gävle, 801 76, Gävle, Sweden

[2]Department of Medical Epidemiology and Biostatistics, Karolinska Institute, Box 281, SE 171 77, Stockholm, Sweden

[*]Corresponding author. E-mail: Li.Yin@ki.se, Phone: +46-8-52486187, Fax: +46-8- 311101

Summary: In observational studies for the interaction between exposures on dichotomous outcome of a population, one usually uses one parameter of a regression model to describe the interaction, leading to one measure of the interaction. In this article, we use the conditional risk of outcome given exposures and covariates to describe the interaction and obtain five different measures for the interaction in observational studies, i.e. difference between the marginal risk differences, ratio of the marginal risk ratios, ratio of the marginal odds ratios, ratio of the conditional risk ratios, and ratio of the conditional odds ratios. By using only one regression model for the conditional risk of outcome given exposures and covariates, we obtain the maximum-likelihood estimates of all these measures. By generating approximate distributions of the maximum-likelihood estimates of these measures, we obtain interval estimates of these measures. The method is presented by studying the interaction between a therapy and the environment on eradication of *Helicobacter pylori* among Vietnamese children.

## 1. Introduction

The interaction between exposures on an outcome refers to the situation where the effect of one exposure on the outcome depends on the assignment of another exposure [21, 22, 24]. Suppose a hypothetical randomized trial in which one could assign two exposures $z_1$ and $z_2$ and wished to investigate the effect of two exposures $z_1$ and $z_2$ on an outcome $y$ of certain population, where the variables $z_1$, $z_2$ and $y$ are all dichotomous, i.e. $z_1 = 0, 1$, $z_2 = 0, 1$ and $y = 0, 1$. In the randomized trial, covariates are essentially unassociated with exposures ($z_1$, $z_2$) and thus are not confounders. In this case, one may use the marginal risk $\text{pr}(y = 1 \mid z_1, z_2)$ to measure the effect of $z_1$ and then the interaction between $z_1$ and $z_2$. Here the marginal risk $\text{pr}(y = 1 \mid z_1, z_2)$ is marginal with respect to covariates and thus conditional on exposures ($z_1$, $z_2$) only.

One may measure the effect of $z_1$ by the marginal risk difference

$$\text{MRD}(z_2) = \text{pr}(y = 1 \mid z_1 = 1, z_2) - \text{pr}(y = 1 \mid z_1 = 0, z_2)$$

and the interaction by difference between the marginal risk differences

$$\text{DMRD} = \text{MRD}(z_2{=}1) - \text{MRD}(z_2{=}0).$$

One may also measure the effect of $z_1$ by the marginal risk ratio

$$\text{MRR}(z_2) = \text{pr}(y = 1 \mid z_1 = 1, z_2)/\text{pr}(y = 1 \mid z_1 = 0, z_2)$$

and the interaction by ratio of the marginal risk ratios

$$\text{RMRR} = \text{MRR}(z_2{=}1)/\text{MRR}(z_2{=}0).$$

One may also measure the effect of $z_1$ by the marginal odds ratio

$$\text{MOR}(z_2) = \frac{\text{pr}(y = 1 \mid z_1 = 1, z_2)/\{1 - \text{pr}(y = 1 \mid z_1 = 1, z_2)\}}{\text{pr}(y = 1 \mid z_1 = 0, z_2)/\{1 - \text{pr}(y = 1 \mid z_1 = 0, z_2)\}}$$

and the interaction by ratio of the conditional odds ratios

$$\text{RMOR} = \text{MOR}(z_2{=}1)/\text{MOR}(z_2{=}0).$$

Oftentimes, DMRD is called the biological interaction because its expression in terms of probability differences has a close relationship with some well-known classification of biological mechanisms [7, 21]. However, the other two measures RMRR and RMOR of the interaction are also commonly used, have causal interpretation in the framework of Rubin causal model [19, 20, 23], and reflect different aspects of the underlying the interaction [24].

Now suppose an observational study in which one investigates the effect of the exposure ($z_1$, $z_2$) on the outcome $y$ of the population above. In the observational study, covariates may be associated with both the outcome $y$ and the exposure ($z_1$, $z_2$) and therefore are confounders. For illustrative purpose, we consider the case of one cofounder $x$. To adjust for $x$ in estimation of the effect of ($z_1$, $z_2$), one needs to model the conditional risk $\text{pr}(y = 1|z_1, z_2, x)\}$ given not only ($z_1$, $z_2$) but also $x$. When sample size is small, one often obtains a simple model for $\text{pr}(y = 1|z_1, z_2, x)$, for instance,

$$\text{g}\{\text{pr}(y = 1|z_1, z_2, x)\} = \alpha + \beta_1 z_1 + \beta_2 z_2 + \beta_3(z_1 * z_2) + \theta_1 x$$

where $\text{g}(.)$is a link function. The model contains one term for the product $z_1 * z_2$, but no terms for the products between $x$ and ($z_1$, $z_2$). The parameter $\beta_3$ is called the statistical interaction which describes deviation of this model from the main effect model containing no term for any product between variables [7, 21, 22].

If there are no unmeasured confounders than the confounder $x$, then $\beta_3$ or a function of $\beta_3$ has causal interpretation and measures the interaction on any stratum $x = 0, 1$ and thus on the population. With the identity link function $\text{g}(p) = p$, the parameter $\beta_3$ is difference between the conditional risk differences (DCRD), which is also equal to difference between the marginal risk differences (DMRD) because risk differences are collapsible. With the log link function $\text{g}(p) = \log(p)$, the exponential $\exp(\beta_3)$ is ratio of the conditional risk ratios (RCRR), which is also equal to ratio of the marginal risk ratios (RMRR). With the logistic link function $\text{g}(p) = \text{logit}(p)$, the exponential $\exp(\beta_3)$ is ratio of the conditional odds ratios (RCOR), but which is not equal to ratio of the marginal odds ratios (RMOR) due to non-collapsibility of odds ratios. See, for instance, [5, 10, 12, 14] for collapsibility of measures of the exposure effect. In the method above, one first has a model and then uses a parameter of the model to measure the interaction. As a consequence, one uses one model to estimate only one measure of the interaction.

When sample size is large, the model inevitably becomes complex, for instance,

$$\text{g}\{\text{pr}(y = 1|z_1, z_2, x)\} = \alpha + \beta_1 z_1 + \beta_2 z_2 + \beta_3(z_1 * z_2) + \theta_1 x + \theta_2(z_1 * x) + \theta_3(z_2 * x)$$

which contains, in addition to one term for the product $z_1 * z_2$, two terms for the products $z_1 * x$ and $z_2 * x$. The parameter $\beta_3$ or a function of $\beta_3$ still measures the interaction. With a linear function, $\beta_3$ is DCRD which is equal to DMRD. With a log link function, $\exp(\beta_3)$ is RCRR but no longer equal to RMRR. With a logistic link function, $\exp(\beta_3)$ is RCOR but still not equal to RMOR. In practical researches, one usually omits the last two terms in the model despite $\theta_2 \neq 0$ and $\theta_3 \neq 0$, leading to a mis-specified model for $\text{pr}(y = 1|z_1, z_2, x)$. Then the parameter $\beta_3$ does not measure the interaction. Furthermore, the model may even contain a triple-product term $\theta_4(z_1 * z_2 * x)$. In this case, the parameter $\beta_3$ measures the interaction on stratum $x = 0$ but not on the population. If one omits this triple-product term despite $\theta_4 \neq 0$, one obtains a mis-specified model for $\text{pr}(y = 1|z_1, z_2, x)$, but then, $\beta_3$ does not measure the interaction on stratum $x = 0$ nor on the population. With these mis-specified models, it would not be possible to have consistent estimate of any measure of the interaction.

In this article, we measure and estimate the interaction between exposures in two separate steps. In the measuring step, we express all the five measures of the interaction, DMRD, RMRR, RMOR, RCRR and RCOR, in terms of the conditional risk $\text{pr}(y = 1|z_1, z_2, \boldsymbol{x})$, rather than model parameters, where $\boldsymbol{x}$ is a vector of covariates. In the estimating step, we estimate DMRD, RMRR, RMOR, RCRR and RCOR through $\text{pr}(y = 1|z_1, z_2, \boldsymbol{x})$, which can be described by any model. As a result, we can use one model in one study to estimate all these measures.

Our method of measuring and estimating the interaction is extension of a well-known method of measuring and estimating the exposure effect. In the latter method, one expresses a measure of the exposure effect in terms of $\text{pr}(y = 1|z_1, z_2, \boldsymbol{x})$ and then estimates the measure through $\text{pr}(y = 1|z_1, z_2, \boldsymbol{x})$ [1,6, 8, 13]. The difficulty with this method is interval estimation for the measure of the exposure effect. The common method for the interval estimation is the parametric or non-parametric bootstrap method, but it is highly difficult to correct the finite-sample bias arising from the bootstrap sampling with dichotomous outcome [3, 4, 9]. To avoid complexity of the bootstrap method for interval estimation of a measure of the interaction, we use distribution of the ML estimates of the model parameters to obtain distribution of the ML estimate of the measure and then the corresponding confidence interval.

We are going to present the method through an observational study embedded in a randomized trial, in which we investigated the interaction between a therapy and the environment on eradication of *Helicobacter pylori* among Vietnamese children.

## 2. The interaction between a therapy and the environment on eradication of *Helicobacter pylori* among Vietnamese children

### 2.1 Medical background and the data

In a randomized trial [18], researchers studied two triple therapies – (lansoprazole, amoxicillin, metronidazole) and (lansoprazole, amoxicillin, clarithromycin) – for their abilities to eradicate *Helicobacter pylori* among Vietnamese children. From several children hospitals in Hanoi, restricting body weight to a range between 13 kg and 45 kg, a sample was collected between May 2005 and January 2006. In an observational substudy embedded in this randomized trial, they focused on one treatment arm of the triple therapy (lansoprazole, amoxicillin, metronidazole). They analyzed the effect of high versus low doses of the therapy and the effect of the environment encoded by geographic areas in which these children lived. In this article, we study how the environment interacted with the therapy on eradication of *Helicobacter pylori* among Vietnamese children. The treatment arm comprised 109 patients.

The therapy eradicated *Helicobacter pylori* through systemic circulation, so the researchers assigned the therapy to the children according to their body weights. According to the pediatric procedure, children with 13 kg < body weight < 23 kg received the therapy once daily and those with 23 kg ≤ weight < 45 kg twice daily. Of medical relevance was dose in unit body weight. Following the earlier report on the same data [18], we categorized the children into receiving low versus high doses of the therapy by taking the middle point of each weight category as the threshold.

Hence children with 13 kg < body weight ≤18 kg in the first weight category and 23 kg ≤ body weight < 34 kg in the second weight category were considered as receiving the high dose ($z_1$ = 1) of the therapy whereas those with 18 kg < body weight < 23 kg in the first weight category and 34 kg ≤ body weight < 45 kg in the second weight category as receiving the low dose ($z_1$ = 0). Let $z_2$ = 1 indicate the rural environment and $z_2$ = 0 the urban environment.

Then the exposures in this study were $\boldsymbol{z} = (z_1, z_2)$ = (therapy, environment). The outcome was successful ($y$ = 1) versus unsuccessful ($y$ = 0) eradication of *Helicobacter pylori*. In this article, we estimate the interaction between exposures $z_1$ and $z_2$ on the outcome $y$ = 1 among Vietnamese children.

In the context of this study, body weight was synonymous to exposure $z_1$ and thus was not an independent covariate. The covariate age ($x_1$) was associated with body weight and thus with the assignment of exposure $z_1$. Furthermore, age was known to have an effect on the outcome $y$ and thus was a confounding covariate. In addition to age, other possible confounding covariates were recorded, which were gender ($x_2$) and antibiotic resistance to metronidazole ($x_3$). Age was categorized into younger ($x_1$ = 1) *versus* older ($x_1$ = 0) than 9 years. Let $x_2$ = 1 indicate female and $x_2$ = 0 male of the gender. Antibiotic resistance to metronidazole was categorized into non-resistant ($x_3$ = 1) *versus* resistant ($x_3$ = 0). Let $\boldsymbol{x} = (x_1, x_2, x_3)$ be the set of all the recorded covariates. The data of this study is given in Table 1.

### 2.2 Supposition for interaction of causal interpretation

In this article, we aim at the interaction of causal interpretation, that is, we compare the potential outcomes $y(\boldsymbol{z})$ of the patients in the population under different exposures $\boldsymbol{z} = (z_1, z_2)$ = (0, 0), (0, 1), (1, 0), (1, 1). Because it is only possible to observe potential outcome of a patient under one value of the exposure $\boldsymbol{z}$, the following supposition is needed to allow for estimation of the interaction [19, 20, 23].

**Supposition** The assignment of exposures $\boldsymbol{z}$ is strongly ignorable given the covariates $\boldsymbol{x}$ and the probability $\mathrm{pr}(\boldsymbol{z}|\boldsymbol{x})$ of $\boldsymbol{z}$ given $\boldsymbol{x}$ is larger than zero**.** Therefore the risk $\mathrm{pr}\{y(\boldsymbol{z})|\boldsymbol{x}\}$ of the potential outcome $y(\boldsymbol{z})$ in stratum $\boldsymbol{x}$ under exposure $\boldsymbol{z}$ is equal to the risk $\mathrm{pr}(y \mid \boldsymbol{z}, \boldsymbol{x})$ of the observable outcome $y$ in stratum $(\boldsymbol{x}, \boldsymbol{z})$, that is, $\mathrm{pr}\{y(\boldsymbol{z})|\boldsymbol{x}\} = \mathrm{pr}(y \mid \boldsymbol{z}, \boldsymbol{x})$.

The supposition is also called the assumption of no unmeasured confounding covariates given the covariates $\boldsymbol{x}$ [11, 21]; see [24] for examples of the supposition and their causal directed acyclic graphs.

## 3. Measuring interaction on population

### 3.1 Conditional measures of interaction on population

The conditional odds $\mathrm{CO}(\boldsymbol{z}; \boldsymbol{x})$ of the potential outcome $y(\boldsymbol{z}) = 1$ in stratum $\boldsymbol{x} = (x_1, x_2, x_3)$ under exposure $\boldsymbol{z} = (z_1, z_2)$ is

$$\mathrm{CO}(\boldsymbol{z};\, \boldsymbol{x}) = \frac{\mathrm{pr}\{y(\boldsymbol{z}) = 1|\boldsymbol{x}\}}{\mathrm{pr}\{y(\boldsymbol{z}) = 0|\boldsymbol{x}\}}, \tag{1}$$

where $\mathrm{pr}\{y(\boldsymbol{z}) = 0|\boldsymbol{x}\} = 1 - \mathrm{pr}\{y(\boldsymbol{z}) = 1|\boldsymbol{x}\}$.

The conditional odds ratio is

$$\mathrm{COR}(z_2;\ \boldsymbol{x}) = \frac{\mathrm{CO}(z_1 = 1,\ z_2;\ \boldsymbol{x})}{\mathrm{CO}(z_1 = 0,\ z_2;\ \boldsymbol{x})}, \tag{2}$$

which measures the effect of the therapy $z_1$ on stratum $\boldsymbol{x}$ in which each patient were assigned environment $z_2$. Then the ratio of the conditional odds ratios is

$$\mathrm{RCOR}(\boldsymbol{x}) = \frac{\mathrm{COR}(z_2 = 1;\ \boldsymbol{x})}{\mathrm{COR}(z_2 = 0;\ \boldsymbol{x})}, \tag{3}$$

which measures the interaction between the exposures $z_1$ and $z_2$ on stratum $\boldsymbol{x}$. Taking the average of RCOR($\boldsymbol{x}$) over $\boldsymbol{x}$ of the population, we obtain the ratio of the conditional odds ratio on the population

$$\mathrm{RCOR} = \sum_{\boldsymbol{x}} \mathrm{pr}(\boldsymbol{x})\,\mathrm{RCOR}(\boldsymbol{x}), \tag{4}$$

which measures the interaction on the population on the conditional odds ratio scale. In particular, if RCOR($\boldsymbol{x}$) takes the same value for all $\boldsymbol{x}$ values, then RCOR($\boldsymbol{x}$) = RCOR. Under Supposition in section 2.2, we have $\mathrm{pr}\{y(\boldsymbol{z})|\boldsymbol{x}\} = \mathrm{pr}(y \mid \boldsymbol{z}, \boldsymbol{x})$. Inserting this equality into (1) and then using (2)-(4), we obtain

$$\mathrm{RCOR} = \sum_{\boldsymbol{x}} \mathrm{pr}(\boldsymbol{x}) \frac{\left\{\dfrac{\mathrm{pr}(y = 1 \mid z_1 = 1,\ z_2 = 1,\ \boldsymbol{x})\mathrm{pr}(y = 0 \mid z_1 = 0,\ z_2 = 1,\ \boldsymbol{x})}{\mathrm{pr}(y = 0 \mid z_1 = 1,\ z_2 = 1,\ \boldsymbol{x})\mathrm{pr}(y = 1 \mid z_1 = 0,\ z_2 = 1,\ \boldsymbol{x})}\right\}}{\left\{\dfrac{\mathrm{pr}(y = 1 \mid z_1 = 1,\ z_2 = 0,\ \boldsymbol{x})\mathrm{pr}(y = 0 \mid z_1 = 0,\ z_2 = 0,\ \boldsymbol{x})}{\mathrm{pr}(y = 0 \mid z_1 = 1,\ z_2 = 0,\ \boldsymbol{x})\mathrm{pr}(y = 1 \mid z_1 = 0,\ z_2 = 0,\ \boldsymbol{x})}\right\}}. \tag{5}$$

The conditional risk ratio is

$$\mathrm{CRR}(z_2;\ \boldsymbol{x}) = \frac{\mathrm{pr}\{y(z_1 = 1, z_2) = 1|\boldsymbol{x}\}}{\mathrm{pr}\{y(z_1 = 0, z_2) = 1|\boldsymbol{x}\}}, \tag{6}$$

which measures the effect of $z_1$ on stratum $\boldsymbol{x}$ in which each patient were assigned $z_2$. The ratio of the conditional risk ratios is

$$\mathrm{RCRR}(\boldsymbol{x}) = \frac{\mathrm{CRR}(z_2 = 1;\ \boldsymbol{x})}{\mathrm{CRR}(z_2 = 0;\ \boldsymbol{x})}, \tag{7}$$

which measures the interaction on stratum $\boldsymbol{x}$. Taking the average of RCRR($\boldsymbol{x}$) over $\boldsymbol{x}$ of the population, we obtain the ratio of the conditional risk ratios on the population

$$\mathrm{RCRR} = \sum_{\boldsymbol{x}} \mathrm{pr}(\boldsymbol{x})\,\mathrm{RCRR}(\boldsymbol{x}), \tag{8}$$

which measures the interaction on the population on the conditional risk ratio scale. In particular, if RCRR($\boldsymbol{x}$) takes the same value for all $\boldsymbol{x}$ values, then RCRR($\boldsymbol{x}$) = RCRR. Inserting the equality $\mathrm{pr}\{y(\boldsymbol{z})|\boldsymbol{x}\} = \mathrm{pr}(y \mid \boldsymbol{z}, \boldsymbol{x})$ into (6) and then using (6)-(8), we obtain

$$\mathrm{RCRR} = \sum_{\boldsymbol{x}} \mathrm{pr}(\boldsymbol{x}) \frac{\left\{\dfrac{\mathrm{pr}(y = 1 \mid z_1 = 1,\ z_2 = 1,\ \boldsymbol{x})}{\mathrm{pr}(y = 1 \mid z_1 = 0,\ z_2 = 1,\ \boldsymbol{x})}\right\}}{\left\{\dfrac{\mathrm{pr}(y = 1 \mid z_1 = 1,\ z_2 = 0,\ \boldsymbol{x})}{\mathrm{pr}(y = 1 \mid z_1 = 0,\ z_2 = 0,\ \boldsymbol{x})}\right\}}. \tag{9}$$

We may also use the conditional risk difference

$$\mathrm{CRD}(z_2;\ \boldsymbol{x}) = \mathrm{pr}\{y(z_1 = 1, z_2) = 1|\boldsymbol{x}\} - \mathrm{pr}\{y(z_1 = 0, z_2) = 1|\boldsymbol{x}\} \tag{10}$$

to measure the effect of $z_1$ on stratum $\boldsymbol{x}$ in which each patient were assigned $z_2$, and the difference between the conditional risk differences

$$\mathrm{DCRD}(\boldsymbol{x}) = \mathrm{CRD}(z_2 = 1;\ \boldsymbol{x}) - \mathrm{CRD}(z_2 = 0;\ \boldsymbol{x}) \tag{11}$$

to measure the interaction on stratum $\boldsymbol{x}$, and the average of DCRD($\boldsymbol{x}$) over $\boldsymbol{x}$ of the population

$$\mathrm{DCRD} = \sum_{\boldsymbol{x}} \mathrm{pr}(\boldsymbol{x})\mathrm{DCRD}(\boldsymbol{x}) \tag{12}$$

to measure the interaction on the population on the conditional risk difference scale. In particular, if DCRD($\boldsymbol{x}$) takes the same value for all $\boldsymbol{x}$ values, then DCRD($\boldsymbol{x}$) = DCRD. Inserting the equality $\mathrm{pr}\{y(\boldsymbol{z})|\boldsymbol{x}\} = \mathrm{pr}(y \mid \boldsymbol{z}, \boldsymbol{x})$ into (10) and then using (10)-(12), we obtain

$$\begin{aligned}\mathrm{DCRD} = \sum_{\boldsymbol{x}} \mathrm{pr}(\boldsymbol{x}) \, [\{&\mathrm{pr}(y = 1 \mid z_1 = 1,\ z_2 = 1, \boldsymbol{x}) - \mathrm{pr}(y = 1 \mid z_1 = 0,\ z_2 = 1, \boldsymbol{x})\} \\ &- \{\mathrm{pr}(y = 1 \mid z_1 = 1,\ z_2 = 0, \boldsymbol{x}) - \mathrm{pr}(y = 1 \mid z_1 = 0,\ z_2 = 0, \boldsymbol{x})\}].\end{aligned} \tag{13}$$

### 3.2 Marginal measures of interaction on population

The population risk is the risk $\mathrm{pr}\{y(\boldsymbol{z}) = 1\}$ of the potential outcome $y(\boldsymbol{z}) = 1$ of the population under exposure $\boldsymbol{z} = (z_1, z_2)$. We rewrite the population risk $\mathrm{pr}\{y(\boldsymbol{z}) = 1\}$ by

$$\mathrm{pr}\{y(\boldsymbol{z}) = 1\} = \sum_{\boldsymbol{x}} \mathrm{pr}\{y(\boldsymbol{z}) = 1|\boldsymbol{x}\} \, \mathrm{pr}(\boldsymbol{x}).$$

Under Supposition in section 2.2, we have $\mathrm{pr}\{y(\boldsymbol{z}) = 1|\boldsymbol{x}\} = \mathrm{pr}(y = 1 \mid \boldsymbol{z}, \boldsymbol{x})$. Therefore we obtain

$$\mathrm{pr}\{y(\boldsymbol{z}) = 1\} = \sum_{\boldsymbol{x}} \mathrm{pr}(y = 1 \mid \boldsymbol{z}, \boldsymbol{x}) \, \mathrm{pr}(\boldsymbol{x}).$$

Let $\mathrm{PR}(y = 1 \mid \boldsymbol{z}) = \sum_{\boldsymbol{x}} \mathrm{pr}(y = 1 \mid \boldsymbol{z}, \boldsymbol{x}) \, \mathrm{pr}(\boldsymbol{x})$, which is called the population-adjusted risk [6, 8, 13]. Then under Supposition we have that the population-adjusted risk is same as the population risk, namely

$$\mathrm{PR}(y = 1 \mid \boldsymbol{z}) = \mathrm{pr}\{y(\boldsymbol{z}) = 1\} = \sum_{\boldsymbol{x}} \mathrm{pr}(y = 1 \mid \boldsymbol{z}, \boldsymbol{x}) \, \mathrm{pr}(\boldsymbol{x}). \tag{14}$$

The marginal odds MO($\boldsymbol{z}$) of the potential outcome $y(\boldsymbol{z}) = 1$ of the population under exposure $\boldsymbol{z}$ is

$$\mathrm{MO}(\boldsymbol{z}) = \frac{\mathrm{pr}\{y(\boldsymbol{z}) = 1\}}{\mathrm{pr}\{y(\boldsymbol{z}) = 0\}}, \tag{15}$$

where $\mathrm{pr}\{y(\boldsymbol{z}) = 0\} = 1 - \mathrm{pr}\{y(\boldsymbol{z}) = 1\}$. The marginal odds ratio is

$$\mathrm{MOR}(z_2) = \frac{\mathrm{MO}(z_1 = 1,\ z_2)}{\mathrm{MO}(z_1 = 0,\ z_2)}, \tag{16}$$

which measures the effect of the therapy $z_1$ on the population in which all patients were assigned environment $z_2$. The ratio of marginal odds ratios is

$$\mathrm{RMOR} = \frac{\mathrm{MOR}(z_2 = 1)}{\mathrm{MOR}(z_2 = 0)}, \tag{17}$$

which measures the interaction between the exposures $z_1$ and $z_2$ on the population on the marginal odds ratio scale. Using formulas (14)-(17), we obtain

$$\mathrm{RMOR} = \frac{\left\{\frac{\mathrm{PR}(y = 1 \mid z_1 = 1, z_2 = 1)\mathrm{PR}(y = 0 \mid z_1 = 0, z_2 = 1)}{\mathrm{PR}(y = 0 \mid z_1 = 1, z_2 = 1)\mathrm{PR}(y = 1 \mid z_1 = 0, z_2 = 1)}\right\}}{\left\{\frac{\mathrm{PR}(y = 1 \mid z_1 = 1, z_2 = 0)\mathrm{PR}(y = 0 \mid z_1 = 0, z_2 = 0)}{\mathrm{PR}(y = 0 \mid z_1 = 1, z_2 = 0)\mathrm{PR}(y = 1 \mid z_1 = 0, z_2 = 0)}\right\}}. \tag{18}$$

The marginal risk ratio is

$$\mathrm{MRR}(z_2) = \frac{\mathrm{pr}\{y(z_1 = 1, z_2) = 1\}}{\mathrm{pr}\{y(z_1 = 0, z_2) = 1\}}, \tag{19}$$

which measures the effect of $z_1$ on the population in which each patient were assigned $z_2$. The ratio of marginal risk ratios is

$$\mathrm{RMRR} = \frac{\mathrm{MRR}(z_2 = 1)}{\mathrm{MRR}(z_2 = 0)}, \tag{20}$$

which measures the interaction on the population on the marginal risk ratio scale. Using (14), (19) and (20), we obtain

$$\mathrm{RMRR} = \frac{\left\{\frac{\mathrm{PR}(y = 1 \mid z_1 = 1,\ z_2 = 1)}{\mathrm{PR}(y = 1 \mid z_1 = 0,\ z_2 = 1)}\right\}}{\left\{\frac{\mathrm{PR}(y = 1 \mid z_1 = 1,\ z_2 = 0)}{\mathrm{PR}(y = 1 \mid z_1 = 0, z_2 = 0)}\right\}}. \tag{21}$$

The marginal risk difference is

$$\mathrm{MRD}(z_2) = \mathrm{pr}\{y(z_1 = 1, z_2) = 1\} - \mathrm{pr}\{y(z_1 = 0, z_2) = 1\}, \tag{22}$$

which measures the effect of $z_1$ on the population in which each patient were assigned $z_2$. The difference of the marginal risk differences is

$$\mathrm{DMRD} = \mathrm{MRD}(z_2 = 1) - \mathrm{MRD}(z_2 = 0), \tag{23}$$

which measures the interaction on the population on the marginal risk difference scale. Using formulas (14), (22) and (23), we obtain

$$\mathrm{DMRD} = \mathrm{PR}(y = 1 \mid z_1 = 1, z_2 = 1) - \mathrm{PR}(y = 1 \mid z_1 = 0, z_2 = 1) \tag{24}$$
$$-\mathrm{PR}(y = 1 \mid z_1 = 1, z_2 = 0) + \mathrm{PR}(y = 1 \mid z_1 = 0, z_2 = 0).$$

In randomized trial, the covariates $\boldsymbol{x}$ are essentially unassociated with the exposure $\boldsymbol{z} = (z_1, z_2)$, i.e. $\mathrm{pr}(\boldsymbol{x} \mid \boldsymbol{z}) = \mathrm{pr}(\boldsymbol{x})$. As a result, formula (14) becomes

$$\mathrm{pr}\{y(\boldsymbol{z}) = 1\} = \mathrm{PR}(y = 1 \mid \boldsymbol{z}) = \mathrm{pr}(y = 1 \mid \boldsymbol{z}),$$

where $\mathrm{pr}(y = 1 \mid \boldsymbol{z})$ is the marginal risk of the outcome $y = 1$ in stratum $\boldsymbol{z}$. In this case, the marginal measures of the interaction can be obtained from the marginal risk $\mathrm{pr}(y = 1 \mid \boldsymbol{z})$ instead of $\mathrm{pr}(y = 1 \mid \boldsymbol{z}, \boldsymbol{x})$. Therefore the marginal measures we have obtained in this subsection are the same as those we described in the introduction for randomized trial. However, in many situations like the example of this article, randomized trial is not possible, and one can only rely upon observational studies and therefore needs to use the population-adjusted risk $\mathrm{PR}(y = 1 \mid \boldsymbol{z})$ to obtain marginal measures of the interaction on population.

### 3.3 Properties of measures of interaction on population

From (4), we see that RCOR compares the potential outcomes $y(\boldsymbol{z})$ of the patients in the population under different exposures $\boldsymbol{z} = (z_1, z_2) = (0, 0), (0, 1), (1, 0), (1, 1)$ and thus has causal interpretation in the framework of Rubin Causal Model (see, for instance, [19, 20, 23]). Similarly, from (8), (17), (20) and (23), we see that RCRR, RMOR, RMRR and DMRD all have causal interpretation.

These measures are symmetric: we obtain the same formulas for them if we switch the order of exposures $z_1$ and $z_2$. We have RCOR ≠ RMOR and RCRR ≠ RMRR because the odds ratio and the risk ratio are not collapsible [5, 10, 12, 14]. We always have DCRD = DMRD because the risk difference is collapsible. Thus we have five different measures of the interaction.

These measures reflect different aspects of the same underlying interaction: the null hypotheses, i.e. RCOR = 1, RCRR = 1, RMOR = 1, RMRR = 1 and DMRD = 0, do not imply one another. The conditional measures RCOR and RCRR describe the interaction on population in terms of the conditional risk $\text{pr}\{y(\boldsymbol{z}) = 1 \mid \boldsymbol{x}\}$, emphasizing the interaction on individual covariate-specific strata of the population. The marginal measures RMOR, RMRR and DMRD express interaction on population in terms of the population risk $\text{pr}\{y(\boldsymbol{z}) = 1\}$, emphasizing the interaction on the population as a whole.

According to formulas (5) and (9), the conditional measures RCOR and RCRR can be expressed in terms of $\text{pr}(y = 1 \mid \boldsymbol{z}, \boldsymbol{x})$ and $\text{pr}(\boldsymbol{x})$. According to (18), (21) and (24), the marginal measures RMOR, RMRR and DMRD can be expressed in terms of $\text{PR}(y = 1 \mid \boldsymbol{z})$ which in turn can be expressed in terms of $\text{pr}(y = 1 \mid \boldsymbol{z}, \boldsymbol{x})$ and $\text{pr}(\boldsymbol{x})$ according to (14). Therefore all the five measures can be expressed in terms of $\text{pr}(y = 1 \mid \boldsymbol{z}, \boldsymbol{x})$ and $\text{pr}(\boldsymbol{x})$, implying that we can estimate these measures by estimating $\text{pr}(y = 1 \mid \boldsymbol{z}, \boldsymbol{x})$ and $\text{pr}(\boldsymbol{x})$ through observed data. In other words, we can use any model to estimate these measures if the model is correctly specified for $\text{pr}(y = 1 \mid \boldsymbol{z}, \boldsymbol{x})$.

## 4. Estimating interaction on population

### 4.1 Regression model

By fitting the data introduced in Section 2 to a logistic model, we obtain the following model for the risk $\text{pr}(y = 1 \mid \boldsymbol{z}, \boldsymbol{x})$ of $y = 1$ in stratum $(\boldsymbol{z}, \boldsymbol{x})$

$$\text{Log}\left\{\frac{\text{pr}(y = 1 \mid \boldsymbol{z}, \boldsymbol{x})}{1 - \text{pr}(y = 1 \mid \boldsymbol{z}, \boldsymbol{x})}\right\} =$$

$$\alpha + \beta_1 z_1 + \beta_2 z_2 + \beta_3 (z_1 * z_2) + \theta_1 x_1 + \theta_2 x_2 + \theta_3 x_3 + \theta_4 (z_1 * x_2). \qquad (25)$$

In this model, we include, in addition to one term for the product $z_1 * z_2$, another term for the product $z_1 * x_2$, because of the small p-value, 0.10, for the likelihood ratio-based significance test of $\theta_4 = 0$. Let $\pi = (\alpha, \beta_1, \beta_2, \beta_3, \theta_1, \theta_2, \theta_3, \theta_4)$ be the set of all model parameters. The ML estimate $\hat{\pi} = (\hat{\alpha}, \hat{\beta}_1, \hat{\beta}_2, \hat{\beta}_3, \hat{\theta}_1, \hat{\theta}_2, \hat{\theta}_3, \hat{\theta}_4)$ and its approximate covariance matrix $\hat{\Sigma}$ (i.e. the inverse of the observed information) are given in Table 2.

Combining model (25) with formula (9), we have that $\exp(\beta_3)$ is equal to RCOR. In addition to RCOR, we are also interested in other aspects of the interaction on the population described by RCRR, RMOR, RMRR and DMRD, which are not functions of $\beta_3$. In particular, since the outcome $y$ is common as seen in Table 1, the conditional odds ratios are poor approximations of the conditional risk ratios, so we cannot convert conditional odds ratios into other measures of interaction. If we use a log-linear model

$$\log\{\text{pr}(y = 1 \mid \boldsymbol{z}, \boldsymbol{x})\} =$$

$$\alpha + \beta_1 z_1 + \beta_2 z_2 + \beta_3(z_1 * z_2) + \theta_1 x_1 + \theta_2 x_2 + \theta_3 x_3 + \theta_4(z_1 * x_2)$$

then exp($\beta_3$) is equal to RCRR according to (9), but the other measures are not functions of $\beta_3$ of this log-linear model. If we use a linear model

$$\mathrm{pr}(y = 1 \mid \boldsymbol{z}, \boldsymbol{x}) =$$

$$\alpha + \beta_1 z_1 + \beta_2 z_2 + \beta_3(z_1 * z_2) + \theta_1 x_1 + \theta_2 x_2 + \theta_3 x_3 + \theta_4(z_1 * x_2)$$

then $\beta_3$ is equal to DCRD according to (13), but the other measures are not functions of $\beta_3$ of this linear model. Furthermore, the ML estimations of the parameters in the log-linear and linear models do not converge even in the absence of the term for the product $z_1 * x_2$. Robust methods could improve the convergence but need additional assumptions such as the Poisson distribution for the frequency of $y$ = 1 (see, for instance, [2]). Consequently, we are not able to obtain the ML estimates of RCRR and DMRD by these two models without additional assumptions.

If the term for the product $z_1 * x_2$ in model (25) is omitted despite $\theta_4 \neq 0$, then we obtain a mis-specified model for the conditional risk pr($y$ = 1 | $\boldsymbol{z}$, $\boldsymbol{x}$)

$$\mathrm{Log}\left\{\frac{\mathrm{pr}(y = 1 \mid \boldsymbol{z}, \boldsymbol{x})}{1 - \mathrm{pr}(y = 1 \mid \boldsymbol{z}, \boldsymbol{x})}\right\} = \alpha + \beta_1 z_1 + \beta_2 z_2 + \beta_3(z_1 * z_2) + \theta_1 x_1 + \theta_2 x_2 + \theta_3 x_3. \quad (26)$$

This model may lead to biased estimate of pr($y$ = 1 | $\boldsymbol{z}$, $\boldsymbol{x}$) and thus biased estimates of RCOR and RCRR according to formulas (5) and (9). Furthermore, this model may lead to mis-specification of PR($y$ = 1 | $\boldsymbol{z}$) according to formula (14) and thus to biased estimates of RMOR, RMRR and DMRD according to formulas (18), (21) and (24).

In the rest of this section, we are going to use models (25) and (26) separately to estimate RCOR, RCRR, DMRD, RMOR and RMRR based on the data introduced earlier.

## 4.2 Conditional measures of interaction on population

First we describe the procedure of obtaining the ML estimates of the conditional measures RCOR and RCRR based on model (25). In model (25) replacing the parameters $\pi = (\alpha, \beta_1, \beta_2, \beta_3, \theta_1, \theta_2, \theta_3, \theta_4)$ by the ML estimates $\hat{\pi} = (\hat{\alpha}, \hat{\beta}_1, \hat{\beta}_2, \hat{\beta}_3, \hat{\theta}_1, \hat{\theta}_2, \hat{\theta}_3, \hat{\theta}_4)$ given by Table 2, we obtain the ML estimate $\widehat{\mathrm{pr}}(y = 1 \mid \boldsymbol{z}, \boldsymbol{x})$ of the conditional risk pr($y$ = 1 | $\boldsymbol{z}$, $\boldsymbol{x}$). In formulas (5) and (9) replacing pr($y$ = 1 | $\boldsymbol{z}$, $\boldsymbol{x}$) by $\widehat{\mathrm{pr}}(y = 1 \mid \boldsymbol{z}, \boldsymbol{x})$ and pr($\boldsymbol{x}$) by the proportion of $\boldsymbol{x}$ in the sample, we obtain the ML estimates $\widehat{\mathrm{RCOR}}$ = 8.85 and $\widehat{\mathrm{RCRR}}$ = 1.60, which are presented in Table 3.

To obtain interval estimate of RCOR, we generate approximate distribution of the ML estimate $\widehat{\mathrm{RCOR}}$. Normal distribution is poor approximation to the distribution of $\widehat{\mathrm{RCOR}}$, because RCOR ranges from 0 to +∞. We have similar situations for the other measures of the interaction. On the other hand, the model parameters $\pi = (\alpha, \beta_1, \beta_2, \beta_3, \theta_1, \theta_2, \theta_3, \theta_4)$ range from −∞ to +∞, so normal distribution is good approximation to the distribution of the ML estimate $\hat{\pi} = (\hat{\alpha}, \hat{\beta}_1, \hat{\beta}_2, \hat{\beta}_3, \hat{\theta}_1, \hat{\theta}_2, \hat{\theta}_3, \hat{\theta}_4)$ (see, for instance, Lindsey, 1996). Let $p$ be a random variable which follows the normal distribution $\mathrm{N}(\hat{\pi}, \hat{\Sigma})$, namely, $p \sim \mathrm{N}(\hat{\pi}, \hat{\Sigma})$, where $\hat{\pi}$ and $\hat{\Sigma}$ in $\mathrm{N}(\hat{\pi}, \hat{\Sigma})$ are given in Table 2. We are going to use this normal distribution to construct approximate distributions of $\widehat{\mathrm{RCOR}}$ and $\widehat{\mathrm{RCRR}}$ in this subsection and those of $\widehat{\mathrm{RMOR}}$, $\widehat{\mathrm{RMRR}}$ and $\widehat{\mathrm{DMRD}}$ in the next subsection.

Now we describe the iteration procedure of generating approximate distribution of $\widehat{\mathrm{RCOR}}$. First, we draw $p$ from the normal distribution $\mathrm{N}(\hat{\pi}, \hat{\Sigma})$ and replace $\pi$ by $p$ in model (25) to obtain a value of $\widehat{\mathrm{pr}}(y = 1 \mid \boldsymbol{z}, \boldsymbol{x})$. Second, we replace $\mathrm{pr}(y = 1 \mid \boldsymbol{z}, \boldsymbol{x})$ by $\widehat{\mathrm{pr}}(y = 1 \mid \boldsymbol{z}, \boldsymbol{x})$ in (5) to obtain a value of $\widehat{\mathrm{RCOR}}$. We iterate the procedure, 1000 times in this article, to obtain an approximate distribution of $\widehat{\text{RCOR}}$. The obtained approximate distribution is used to obtain the confidence interval of RCOR. Similarly, we use the normal distribution $p \sim \text{N}(\hat{\pi}, \hat{\Sigma})$ and model (25) and formula (9) to obtain an approximate distribution of $\widehat{\text{RCRR}}$ and then the confidence interval of RCRR.

The obtained approximate distributions of $\widehat{\text{RCOR}}$ and $\widehat{\text{RCRR}}$ are shown in Figure 1. The 95 % confidence interval of RCOR is (0.66, 127.04) while that of RCRR is (0.74, 4.60). They are also presented in Table 3, together with 50 % confidence intervals of RCOR and RCRR.

Replacing model (25) by model (26) in the above procedure, we obtain the ML estimates and the confidence intervals of RCOR and RCRR based on model (26). The results are presented in Table 3.

### 4.3 Marginal measures of interaction on population

First we describe the procedure of obtaining the ML estimates of the marginal measures RMOR, RMRR and DMRD based on model (25). In model (25) replacing the parameters $\pi = (\alpha, \beta_1, \beta_2, \beta_3, \theta_1, \theta_2, \theta_3, \theta_4)$ by the ML estimates $\hat{\pi} = (\hat{\alpha},\ \hat{\beta}_1,\ \hat{\beta}_2,\ \hat{\beta}_3,\ \hat{\theta}_1,\ \hat{\theta}_2,\ \hat{\theta}_3,\ \hat{\theta}_4)$ given in Table 2, we obtain the ML estimate $\widehat{\text{pr}}(y = 1 \mid \boldsymbol{z},\ \boldsymbol{x})$ of the conditional risk $\text{pr}(y = 1 \mid \boldsymbol{z}, \boldsymbol{x})$. In formula (14) replacing $\text{pr}(y = 1 \mid \boldsymbol{z}, \boldsymbol{x})$ by $\widehat{\text{pr}}(y = 1 \mid \boldsymbol{z},\ \boldsymbol{x})$ and $\text{pr}(\boldsymbol{x})$ by the proportion of $\boldsymbol{x}$ in the sample, we obtain the ML estimate $\widehat{\text{PR}}(y = 1 \mid \boldsymbol{z})$ of the population-adjusted risk $\text{PR}(y = 1 \mid \boldsymbol{z})$. In formulas (18), (21) and (24) replacing $\text{PR}(y = 1 \mid \boldsymbol{z})$ by $\widehat{\text{PR}}(y = 1 \mid \boldsymbol{z})$, we obtain the ML estimates $\widehat{\text{RMOR}}$ = 8.62, $\widehat{\text{RMRR}}$ = 1.58 and $\widehat{\text{RMRD}}$ = 0.34. These estimates are also presented in Table 3.

Now we describe the iteration procedure of generating approximate distributions of $\widehat{\text{RMOR}}$. First, we draw $p$ from $\text{N}(\hat{\pi}, \hat{\Sigma})$ and replace $\pi$ by $p$ in model (25) to get $\widehat{\text{pr}}(y = 1 \mid \boldsymbol{z}, \boldsymbol{x})$. Second, we replace $\text{pr}(y = 1 \mid \boldsymbol{z}, \boldsymbol{x})$ by $\widehat{\text{pr}}(y = 1 \mid \boldsymbol{z}, \boldsymbol{x})$ in (14) to get $\widehat{\text{PR}}(y = 1 \mid \boldsymbol{z})$. Finally, we replace $\text{PR}(y = 1 \mid \boldsymbol{z})$ by $\widehat{\text{PR}}(y = 1 \mid \boldsymbol{z})$ in formula (18) to get $\widehat{\text{RMOR}}$. We iterate the procedure, 1000 times in this article, to get an approximate distribution of $\widehat{\text{RMOR}}$. The approximate distribution is used to get the confidence interval of RMOR. Similarly, we use the normal distribution $p \sim \text{N}(\hat{\pi}, \hat{\Sigma})$ and formulas (25), (14) and (21) to get an approximate distribution of $\widehat{\text{RMRR}}$ and then the confidence interval of RMRR, and the normal distribution and formulas (25), (14) and (24) to get an approximate distribution of $\widehat{\text{DMRD}}$ and then the confidence interval of DMRD.

The obtained approximate distributions of $\widehat{\text{RMOR}}$, $\widehat{\text{RMRR}}$ and $\widehat{\text{DMRD}}$ are shown in Figure 1. The 95 % confidence interval of RMOR is (0.72, 88.51), that of RMRR is (0.77, 3.40), and that of DMRD is (−0.10, 0.72). These confidence intervals are also presented in Table 3, together with the 50 % confidence intervals of RMOR and RMRR and DMRD.

Replacing model (25) by model (26) in the above procedure, we get the ML estimates and the confidence intervals of RMOR, RMRR and DMRD based on model (26). The results are given in Table 3.

### 4.4 Results

In practical researches, one usually uses model (26), which omits the term for the product $z_1 * x_2$, to measure and estimate the interaction as described in the introduction. In this case, one can only have RCOR. Because the outcome y = 1 is common in this study, one cannot convert RCOR into RCRR or DMRD.

Despite the common outcome y = 1 and the term for the product $z_1 * x_2$, we have used one logistic model (25) to estimate all the measures of interaction, RCOR, RCRR, RMOR, RMRR and DMRD. Figure 1 shows the approximate distributions of the ML estimates $\widehat{\text{RCOR}}$, $\widehat{\text{RCRR}}$, $\widehat{\text{RMOR}}$, $\widehat{\text{RMRR}}$ and $\widehat{\text{DMRD}}$ based on model (25). Table 3 summarizes the ML estimates and the 50 % and 95 % confidence intervals for the five measures RCOR, RCRR, RMOR, RMRR and DMRD based on model (25) in comparison to the mis-specified model (26).

By comparing the second column with the third column in Table 3, we see that model (26) causes considerable biases to all the five measures. These biases might be exacerbated if more terms for complex products between exposures and covariates were needed to model pr($y$ = 1 | $\boldsymbol{z}$, $\boldsymbol{x}$) as typical of a large sample. Our method can avoid these biases by being able to use complex models.

Model (25) fits the data best and thus provides the most efficient point and interval estimates for these measures of interaction. A clear advantage of using one model in one study to estimate different measures of interaction is that one can avoid different assumptions behind different models, which may complicate interpretations of these measures.

Despite somewhat wide confidence intervals, all the five measures of interaction based on model (25) suggest that the therapy would have higher efficacy of eradicating *Helicobacter pylori* if children lived in the rural environment than if children lived in the urban environment. This finding suggests that urbanization may have contributed to the low efficacy of the therapies containing antibiotics among children in developing countries. To our knowledge, little is seen in the literature about how the environment interacts with therapies on eradication of the bacterium [16, 17].

## 5. Discussion and conclusions

An underlying interaction between exposures on a certain population has different aspects described by different measures. Depending on their specific research subjects, researchers need to know one or several aspects of the interaction. In this article, we have used one model in one study to obtain point and interval estimates for five different measures. The outcome can be common. The model may contain terms for the products between exposures and covariates. The estimation is based on maximum likelihood. The method can be implemented by using any software that generates normal distribution.

In the literature, two methods are available to obtain confidence intervals of various measures of exposure effect, i.e. the normal approximation method and the bootstrap method [1, 6, 8, 13], but which seem less used to obtain confidence intervals for measures of interaction. The normal approximation method is to derive approximate variance of the ML estimate of a measure of exposure effect by the delta method and then use the variance to obtain normal approximation confidence interval of the measure of exposure effect. To obtain normal approximation confidence interval of a measure of interaction, one needs to derive approximate variance of the ML estimate of the measure of interaction, but the derivation is tedious particularly for complex models.

The bootstrap method is to generate bootstrap samples by parametric or non-parametric bootstrap method and then use the bootstrap samples to obtain the bootstrap distribution of a measure of exposure effect and then the bootstrap confidence interval of the measure of exposure effect. However, it is highly difficult to correct the finite-sample bias arising from the bootstrap sampling with dichotomous outcome [3, 4, 9], and the difficulty would be exacerbated for a measure of interaction.

In our method, we generate approximate distribution of the ML estimate of a measure of interaction and use the distribution to obtain the confidence interval for the measure of interaction. This method provides a simple but reliable approach to interval estimation of the measure of interaction. Although it is beyond the scope of this article, it is of interest to derive the corrected bootstrap confidence interval and the normal approximation confidence interval and compare them with the maximum-likelihood-based confidence interval obtained in this article.

**Table 1** Successful eradications / total patients stratified by the covariates $x_1$ (age), $x_2$ (gender), and $x_3$ (antibiotic resistance) for the treatment ($z_1$, $z_2$) (therapy, environment).

| Covariates | | | Treatment $(\boldsymbol{z_1}, \boldsymbol{z_2})$ | | | |
|---|---|---|---|---|---|---|
| $\boldsymbol{x_1}$ | $x_2$ | $x_3$ | (0, 0) | (0, 1) | (1, 0) | (1, 1) |
| 0 | 0 | 0 | 3/4 | 6/7 | 0/3 | 8/8 |
| 0 | 0 | 1 | 3/4 | | 1/1 | 1/1 |
| 0 | 1 | 0 | 1/5 | 3/4 | 2/3 | 3/3 |
| 0 | 1 | 1 | 2/4 | | 1/2 | 1/1 |
| 1 | 0 | 0 | 1/2 | 1/1 | 3/8 | 6/6 |
| 1 | 0 | 1 | 3/3 | 0/1 | 1/1 | 3/3 |
| 1 | 1 | 0 | 2/7 | 0/2 | 5/8 | 0/2 |
| 1 | 1 | 1 | 1/2 | 0/3 | 6/9 | 1/1 |

**Table 2** ML estimates and its approximate covariance matrix for parameters of the model (25)*

| Parameters | $\alpha$ | $\beta_1$ | $\beta_2$ | $\beta_3$ | $\theta_1$ | $\theta_2$ | $\theta_3$ | $\theta_4$ |
|---|---|---|---|---|---|---|---|---|
| Estimates | 1.19 | −0.87 | 0.10 | 2.18 | −0.57 | −1.82 | 0.55 | 1.96 |
| Covariance matrix | | | | | | | | |
| $\alpha$ | 0.64 | −0.53 | −0.32 | 0.29 | −0.12 | −0.42 | −0.13 | 0.47 |
| $\beta_1$ | −0.53 | 1.06 | 0.31 | −0.67 | −0.14 | 0.45 | 0.10 | −0.89 |
| $\beta_2$ | −0.32 | 0.31 | 0.72 | −0.71 | 0.00 | 0.05 | 0.06 | −0.07 |
| $\beta_3$ | 0.29 | −0.67 | −0.71 | 1.86 | 0.05 | −0.05 | −0.04 | 0.32 |
| $\theta_1$ | −0.12 | −0.14 | 0.00 | 0.05 | 0.37 | −0.04 | −0.05 | 0.03 |
| $\theta_2$ | −0.42 | 0.45 | 0.05 | −0.05 | −0.04 | 0.69 | −0.01 | −0.69 |
| $\theta_3$ | −0.13 | 0.10 | 0.06 | −0.04 | −0.05 | −0.01 | 0.39 | −0.10 |
| $\theta_4$ | 0.47 | −0.89 | −0.07 | 0.32 | 0.03 | −0.69 | −0.10 | 1.40 |

*The covariance matrix is obtained with adjustment of over-dispersion in the framework of the ML estimation of a mis-specified model or the generalized estimating equation.

**Table 3** ML estimates and confidence intervals for the five measures of the interaction describing how efficacy of the therapy on eradication of *Helicobacter pylori* would increase if children lived in the rural environment of Vietnam, based on models (25) and (26).

| Interaction | ML estimates | | 50 % and 95 % confidence intervals | |
|---|---|---|---|---|
| | Model (25) | Model (26) | Model (25) | Model (26) |
| RCOR | 8.85 | 6.05 | (0.66, 3.61, 21.95, 127.04) | (0.40, 2.18, 13.89, 72.00) |
| RCRR | 1.60 | 1.47 | (0.74, 1.22, 2.27, 4.60) | (0.60, 1.07, 2.05, 4.18) |
| RMOR | 8.62 | 5.47 | (0.72, 3.50, 15.93, 88.51) | (0.42, 2.04, 11.32, 39.45) |
| RMRR | 1.58 | 1.41 | (0.77, 1.23, 1.90, 3.40) | (0.63, 1.06, 1.80, 3.10) |
| RMRD | 0.34 | 0.27 | (−0.10, 0.18, 0.45, 0.72) | (−0.21, 0.09, 0.39, 0.65) |

**Figure 1** Approximate distributions of the ML estimates -- $\widehat{RCOR}$, $\widehat{RCRR}$, $\widehat{RMOR}$, $\widehat{RMRR}$ and $\widehat{DMRD}$ -- for the five measures of the interaction between the therapy and the environment on the eradication of *Helicobacter* pylori among Vietnamese children.

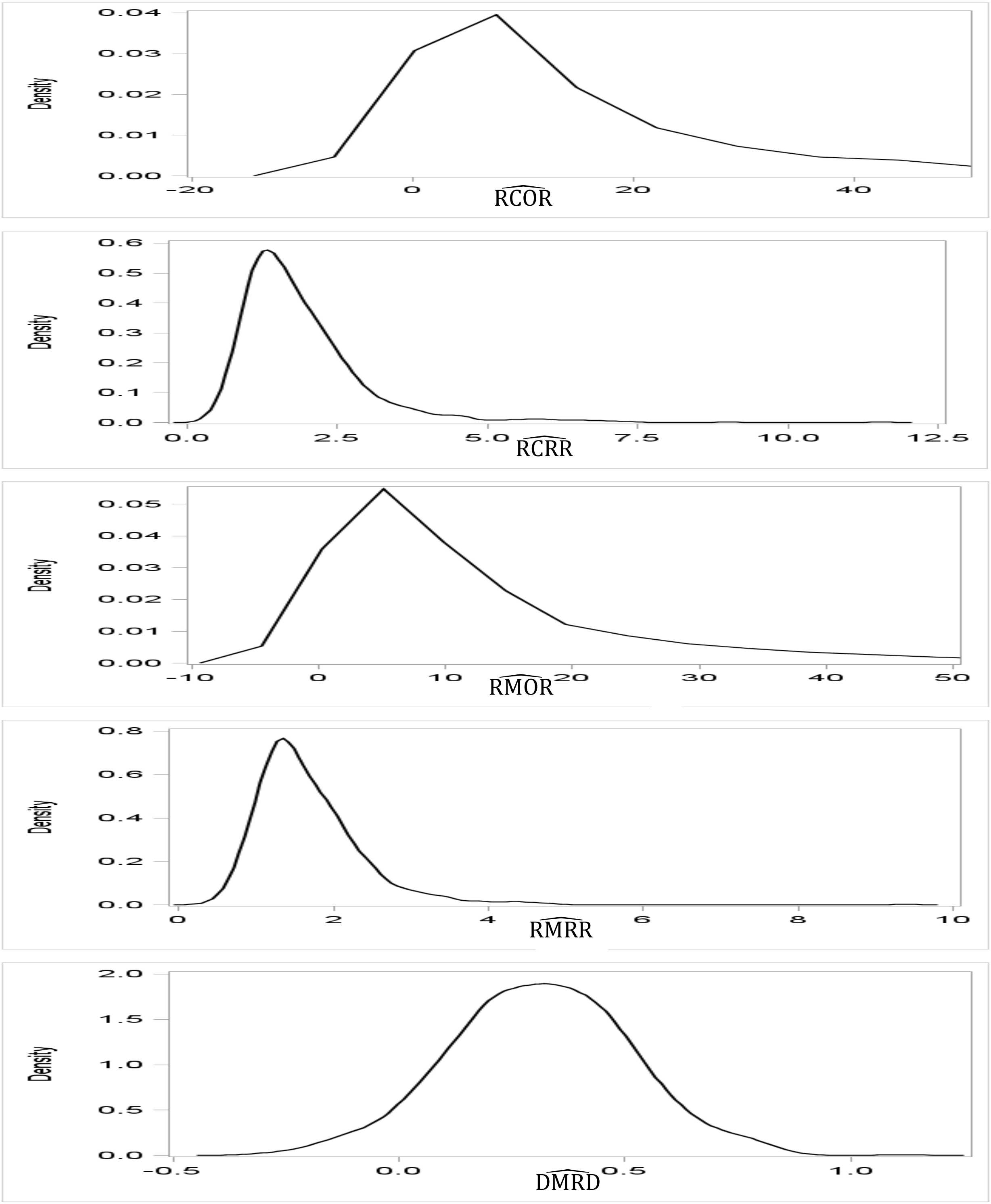